\documentclass[prb,twocolumn,amsmath,amssymb,aps,superscriptaddress,citeautoscript,longbibliography]{revtex4-2}

\usepackage{feynmp}
\usepackage{amsmath}
\usepackage{graphicx}
\usepackage{multirow}
\usepackage{latexsym}
\usepackage{textcomp}
\usepackage{verbatim}
\usepackage{color}
\usepackage{bm}
\usepackage{subfigure}
\usepackage{everysel}
\usepackage{keyval}
\usepackage{ragged2e}
\usepackage{dsfont}
\usepackage{amssymb}
\usepackage{enumitem}
\usepackage{pstricks}
\usepackage{mathtools}
\usepackage{kotex} % Use texlive 2021

\usepackage{braket}
\usepackage{slashed} 
\usepackage{hyperref}        
\hypersetup{
     colorlinks=true,
     citecolor = cyan,    
     linkcolor=magenta,
     }

\usepackage{graphicx}
\usepackage{xcolor}
\usepackage{amsmath}
\usepackage{bbold}
\usepackage{amsfonts}
\usepackage{bbm}
\usepackage{comment}
\usepackage{lettrine}
\usepackage{orcidlink}

\usepackage{placeins}
\newcommand{\beginsupplement}{%
  \setcounter{table}{0}\renewcommand{\thetable}{S\arabic{table}}%
  \setcounter{figure}{0}\renewcommand{\thefigure}{S\arabic{figure}}%
  \setcounter{equation}{0}\renewcommand{\theequation}{S\arabic{equation}}%
}

\newcommand{\osum}{{%
    \setbox0\hbox{\circ}%
    \rlap{\hbox to \wd0{\hss\sum\hss}}\box0
}}

\begin{document}

\title{Comparing the orbital angular momentum and magnetic moment of magnons in the Kagome antiferromagnet with negative spin chirality}
% temporary title

\author{Youngjae Jeon\,\orcidlink{0009-0009-4247-0899}}
%\thanks{Electronic Address: }
\affiliation{Department of Physics, Pohang University of Science and Technology, Pohang 37673, Korea}
\affiliation{Center for Quantum Dynamics of Angular Momentum, Pohang University of Science and Technology, Pohang 37673, Korea}

\author{Jongjun M. Lee\,\orcidlink{0000-0002-9786-1901}}
%\thanks{Electronic Address: }
\affiliation{Department of Physics, Pohang University of Science and Technology, Pohang 37673, Korea}
\affiliation{Center for Quantum Dynamics of Angular Momentum, Pohang University of Science and Technology, Pohang 37673, Korea}
\affiliation{Department of Physics and Quantum Horizons Alberta, University of Alberta, Edmonton, Alberta T6G 2E1, Canada}

\author{Suik Cheon\,\orcidlink{0000-0003-1789-3274}}
%\thanks{Electronic Address: }
\affiliation{Department of Physics, Pohang University of Science and Technology, Pohang 37673, Korea}
\affiliation{Center for Quantum Dynamics of Angular Momentum, Pohang University of Science and Technology, Pohang 37673, Korea}

\author{Hyun-Woo Lee\,\orcidlink{0000-0002-1648-8093}}
%\email{Electronic Address: hwl@postech.ac.kr}
\thanks{Contact author: hwl@postech.ac.kr}
\affiliation{Department of Physics, Pohang University of Science and Technology, Pohang 37673, Korea}
\affiliation{Center for Quantum Dynamics of Angular Momentum, Pohang University of Science and Technology, Pohang 37673, Korea}

%\email[\dagger]{Electronic Address: hwl@postech.ac.kr}

\begin{abstract}
The orbital dynamics of magnons have recently drawn interest due to their potential roles in thermal and orbital transport phenomena in magnetic insulators. In this study, we investigate the orbital magnetic moment (OMM) and orbital angular momentum (OAM) of magnons in a Kagome antiferromagnet with negative vector chirality, focusing on the distinction between thermodynamic and wave-packet-based definitions. We compute the Berry curvature, the OMM, and the OAM in momentum space under an external magnetic field. Our results reveal a quantitative difference between the OMM and OAM, yet their associated Nernst coefficients exhibit similar temperature and field dependence in transport. Our results provide a quantitative comparison between the thermodynamic and wave-packet formulations of magnon orbital dynamics.
\end{abstract}

\date{\today}
\maketitle

\begingroup
\renewcommand\thefootnote{}%
%\footnotetext{* These authors contributed equally to this work.}
%\footnotetext{Electronic Address: hwl@postech.ac.kr}
\addtocounter{footnote}{-1}%
\endgroup

% This is a writing direction. Our aim is the following:
% \begin{itemize}
%     \item What's the novel point or message?
%     \begin{enumerate}
%         \item Difference btw the thermodynamical magnon orbital magnetic moment(MOMM) and modern theoretical magnon orbital angular moment(MOAM).\\
%         \yj{[The name should be discussed. The Nano Letter paper said it as the Magnon Orbital Moment]}
%         \item First calculation of the MOAM in non-collinear material. Especially negative vector chirality Kagome lattice.\\
%         \yj{[check more prior research. Is it really the first calculation?]}
%     \end{enumerate}
%     \item What do I need?
%     \begin{enumerate}
%         \item $E$ vs. $k$ band with or without in-plane DMI.
%         \item Zero spin magnetic moment with or without the in-plane DMI.\\
%         Why does it need? To emphasize the pure MOMM when no magnetic field.
%         \item the MOMM texture vs. the MOAM texture\\
%         Let them know these two are different.\\ \yj{[Does it needed another bands texture? I mean about the lowest band.]}\\ \yj{[How I treat diverge point?]}
%         \item Comparing the MOMM Nernst coefficient and the MOAM Nernst coefficient dependency on temperature.
%         \item Berry curvature needs to be shown?
%         \item Do I need symmetry analysis with or without in-plane DMI?
%     \end{enumerate}
% \end{itemize}

%%%%%%%%%%%%%%%%%%%%%%%%%%%%%%%%%%%%%%%%%%%%%%%%%%%%%%%%
\section{Introduction}
\label{Introduction}
Recently, orbital dynamics in solids have attracted considerable attention~\cite{go2021orbitronics}. In particular, although the orbital angular momentum (OAM) of electrons are quenched due to crystal fields in equilibrium, they still underlie phenomena such as the orbital Hall effect~\cite{bernevig2005orbitronics,tanaka2008intrinsic,kontani2008giant,choi2023observation,lee2025universal} and orbital Edelstein effect~\cite{yoda2015current, chen2018giant, salemi2019orbitally, ding2020harnessing, ding2022observation, lee2024orbital} in nonequilibrium, serving as the origins of their spin counterparts~\cite{sinova2004universal, sinova2015spin, edelstein1990spin, sanchez2013spin, bihlmayer2022rashba}. These orbital characteristics of electrons have led to the emergence of the field of \textit{orbitronics}, which seeks to exploit orbital degrees of freedom for functionalities including orbital torque-induced magnetization control~\cite{hayashi2023observation,han2022orbital,go2023long,sohn2024dyakonov}.
%  Both the atomic orbital degrees of freedom and the itinerant motion of a wave packet contribute to these phenomena, generating the orbital magnetic moment (OMM) via charge dynamics~\cite{chang1996berry,sundaram1999wave,xiao2010berry}.
% JL: Find more proper references for the OHE and OEE. 

For electrons, their OAM is proportional to their orbital magnetic moment (OMM) since they have charge. It is well established that the linear response of the free energy to a magnetic field is directly related, through the gyromagnetic ratio, to the magnetic moment arising from the orbital motion of an electronic wave packet~\cite{chang1996berry,sundaram1999wave,shi2007quantum,xiao2010berry}. Unlike electrons, magnons are chargeless elementary excitations of spin waves~\cite{bloch1930theorie, kittel1948theory}. Therefore, it is not straightforward to relate the magnon OAM, defined through the itinerant motion of a magnon wave packet, to a magnetic moment as in the electronic case.

Recent developments have provided a gauge-invariant formulation of the magnon OAM and revealed that it can exhibit a nontrivial momentum–space texture~\cite{fishman2022orbital, fishman2022exact, fishman2023gauge}. Through the Dzyaloshinskii–Moriya interaction (DMI), this texture can induce an electric polarization~\cite{go2024magnon, tang2025proper, to2025magnon}. While such reports imply a potential coupling of the magnon OAM to the electron change degree of freedom, its relation to the magnon magnetic moment remains unclear. For magnons, their magnetic moment consists of both spin and orbital contributions, with the latter referred to as the magnon OMM. It was reported~\cite{neumann2020orbital} that the magnetic-field-induced change of the free energy of magnetic ordering cannot be completely attributed to the  spin magnetic moment of magnons and the remaining contribution should be attributed to the OMM of magnons. Mathematically, the OMM represents the extent to which the direction of local spin moments changes under an applied magnetic field, and it is a function of crystal momentum as the OAM. These considerations motivate a direct comparison between the magnon OAM and OMM to elucidate their physical characteristics. Establishing their relationship is essential for clarifying the physical meaning of the orbital motion of magnon and its role in transport phenomena.

In this paper, we investigate the relationship between the OMM and OAM of magnons. To this end, we consider a Kagome antiferromagnet with negative vector chirality, where a finite out-of-plane component of the magnon OMM arises~\cite{neumann2020orbital}. The negative vector chirality state hosts a coplanar but noncollinear spin configuration that generates a substantial Berry curvature in the magnon bands in the presence of DMI~\cite{mook2019thermal}. This makes it an ideal platform in which both OAM and OMM responses can coexist. By applying an external magnetic field, we examine how the equilibrium momentum-space textures of the OMM and OAM evolve with the field and compare their respective Nernst coefficients, which quantify the transverse transport induced by a temperature gradient. We find that the equilibrium averages of the OMM and OAM show distinct behavior but their Nernst coefficients exhibit nearly identical field dependencies. This difference originates from the fact that, unlike the OAM, the OMM develops a pronounced peak around the $\Gamma$ point as the field increases. Nervertheless, we find that the Nernst coefficients of the OMM and OAM exhibit nearly identical field dependence, signaling intriguing connection between the OMM and OAM despite the magnons being chargeless. %Our results therefore suggest that, although the equilibrium values of the OMM and OAM can differ significantly, their Nernst responses are expected to display similar trends in a broad class of systems.
% Because the Nernst response is dominated by inter-band effects, whereas the equilibrium averages mainly reflect intra-band contributions, the OMM and OAM yield similar Nernst behaviors despite their contrasting equilibrium distributions.

The paper is organized as follows. In Sec.~\ref{Model and magnetic order}, we introduce the spin model and derive expressions for the canting angle in the presence of an external magnetic field, and comment on the magnetic point group symmetry of the system. In Sec.~\ref{Energy diagram and Berry curvature}, we present the magnon energy spectrum and Berry curvature. Sections~\ref{Magnon orbital magnetic moment} and~\ref{Magnon orbital angular momentum} describe the calculated OMM and OAM textures under an external magnetic field, respectively. In Sec.~\ref{Transport of the magnon}, we formulate the OMM and OAM Nernst coefficients and introduce the corresponding OMM Berry curvature and OAM Berry curvature. In Sec.~\ref{Results and Discussion}, we compare the thermodynamic averages and Nernst responses of the OMM and OAM as functions of temperature and magnetic field. Finally, in Sec .~\ref {Conclusion and discussion}, we conclude with a summary and discussion.

\section{Model and magnetic order}
\label{Model and magnetic order}

% \begin{itemize}
%     \item Spin Hamiltonian\\
%     Spin Hamiltonian with DMI
%     \item{Spin orderings (mean-field analysis)}\\
%     refer to the sign of out-of-plane DMI($D_z$), in-plane DMI($D_p$) effect, and its role
%     \item{Magnon Hamiltonian}\\
%     Should I write the 0-th and 1st order?
%     \item{Magnon energy band}\\
%     with $D_p$ and without it...and referring to the symmetry is good or not?
%     \item{Zero spin expectation value in $z$-axis}\\
%     Let them know the spin expectation value in $z$-axis is zero thus the magnetic moment from spin is zero.
% \end{itemize}

\begin{figure}
    \centering
    \includegraphics[width=\linewidth]{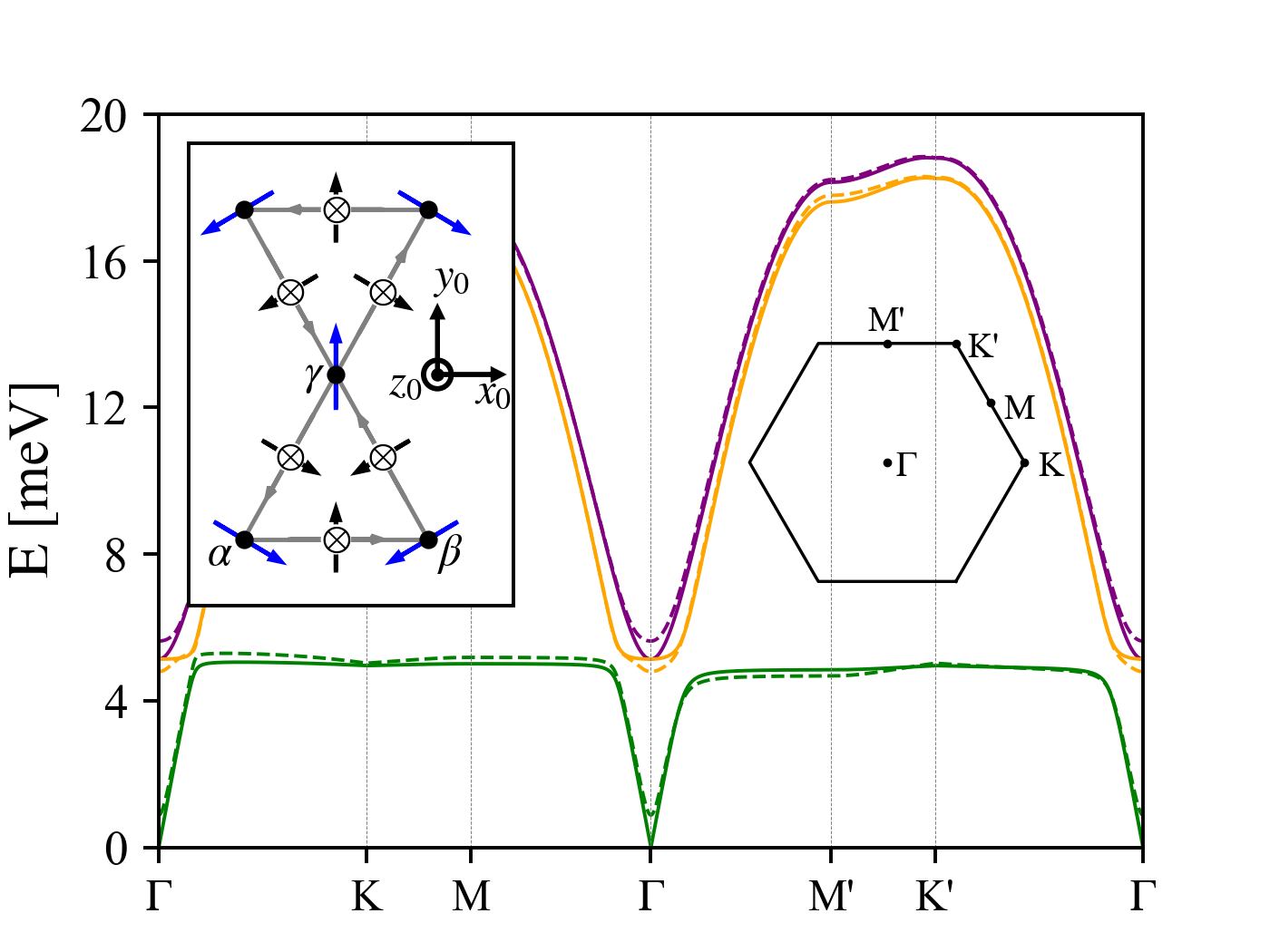}
    \caption{Magnon band structure of the negative vector chirality Kagome lattice. The solid (dashed) lines represent the results in the absence (presence) of a magnetic field $B_{z_0}$. The left inset represent the ground-state spin configuration and the right inset shows the high symmetry point in the first Brillouin zone of the negative vector chirality Kagome lattice. The magnon band structure is obtained from Eq.~(\ref{eq:Hamiltonian}) with $J=3.18 \mbox{ meV},S=5/2, D_{z_0}=-0.062J, D_p=1 \mbox{ meV}, B_{z_0} = 0\mbox{ T } (8\mbox{ T})$ to illustrate the gap opening at the $\Gamma$ point. At $B_{z_0}=8\,\mathrm{T}$, the field-induced gap at $\Gamma$ point is $\epsilon_{1,\Gamma}\simeq0.88\,\mathrm{meV}$, corresponding to the approximately $10.2\,\mathrm{K}$}.
    %\textcolor{red}{1. left inset: magnetic configuration and DMI vectors are denoted. 2. Bands are opened when the external magnetic field is turned on at $\Gamma$ point. 3. right inset: HSP in fBZ. 4. Spins lie in Kagome plane when the external magnetic field is turned off.}
    \label{fig1}
\end{figure}

Here we consider a 2D negative vector chirality Kagome lattice (see inset in Fig.~\ref{fig1}.)
\begin{align}
\hat{H}=\sum_{\left<ij\right>}\left(J\,\hat{\mathbf{S}}_i\cdot\hat{\mathbf{S}}_j+\mathbf{D}_{ij}\cdot\hat{\mathbf{S}}_i\times\hat{\mathbf{S}}_j\right)-g\mu_B B_{z_0}\sum_{i}\hat{S}^{z_0}_i,
\label{eq:Hamiltonian}
\end{align}
where $\mathbf{S}_i$ is spin at site $i$, $J$ is the antiferromagnetic exchange coupling, $\mathbf{D}_{ij}$ denotes the DMI vector associated with $ij$ bond, comprising both in-plane ($D_p$) and out-of-plane ($D_{z_0}$) components, $g$ is the g-factor, $\mu_B$ is the Bohr magneton, and $B_{z_0}$ is the applied magnetic field perpendicular to the Kagome plane. We note that finite $D_{z_0}$ stabilizes the $120^{\circ}$ structure, with negative (positive) $D_{z_0}$ favoring a negative (positive) vector chirality structure~\cite{elhajal2002symmetry}, in which the sign of $D_{z_0}$ is opposite to that in the convention of Ref.~\cite{matan2011dzyaloshinskii, laurell2018magnon}. We adopted the convention that the grey arrows in the inset of Fig.~\ref{fig1} indicate the ordered bond direction $j\to i$ for $\mathbf{D}_{ij}$. In this convention, the $z_0$ component of $\mathbf{D}_{\beta\alpha}, \mathbf{D}_{\gamma\beta}$, and  $\mathbf{D}_{\alpha\gamma}$ is taken to be negative (see inset in Fig.~\ref{fig1}). First we investigate the classical energy to find the relation between local spin direction and $B_{z_0}$. To this end, we consider a spin configuration (left inset in Fig.~\ref{fig1}) with three sublattices $i=\alpha,\beta,\gamma$, whose classical spin directions are given by
\begin{align}
    \mathbf{S}_i=S(-\sin{\theta_i}\cos{\eta},\,\cos{\theta_i}\cos{\eta},\,\sin{\eta}),
\end{align}
where $\theta_i=4\pi/3, 2\pi/3,0$ correspond to the negative vector chirality structure and $\eta$ is the canting angle out of the plane. Using this configuration, the classical energy per spin becomes
\begin{align}
\frac{E_{\mathrm{cl}}(\eta)}{NS^2}=&\frac{J}{2}\left(1-3\cos{2\eta}\right)+\sqrt{3}D_{z_0}\cos^2{\eta}-\frac{g\mu_B B_{z_0}}{S}\sin{\eta},
\end{align}
where $N$ is the number of spins. 
This is an extension of Ref.~\cite{elhajal2002symmetry} by including the Zeeman coupling. A negative $D_{z_0}$ energetically favors the negative vector chirality structure by aligning the vector spin chirality $\mathbf{S}_i\times\mathbf{S}_j$ along the out-of-plane direction, which lowers the DMI energy. By minimizing the classical energy with respect to $\eta$, we obtain
\begin{align}
    \sin{\eta} = \frac{g\mu_B}{S}\frac{B_{z_0}}{6J-2\sqrt{3}D_{z_0}}.
\end{align}
In contrast to the positive vector chirality case~\cite{to2025tunable}, the negative vector chirality Kagome lattice is not tilted by $D_p$; when no magnetic field is applied, the spins lie entirely within the Kagome plane. In this spin structure, a two-fold rotation symmetry about the $x$-axis or mirror symmetry with respect to the $yz$-plane the time-reversal becomes a symmetry when combined with either. These symmetries together form the magnetic point group  $2'/m'$~\cite{mook2019thermal}, which remains preserved regardless of whether an external magnetic field is applied.
%%%%%%%%%%%%%%%%%%%%%%%%%%%%%%%%%%%%%%%%%%%%%%%%%%%%%%%%%%%%%%%%%%%%%%%%%%%%%%%%%%%%%%%%%%%%%%%%%%%%%%%%%%%%%%%%%%%%%%%%%%%%%%%%%%%%%%%%%%%%%%%%%%%
%%%%%%%%%%%%%%%%%%%%%%%%%%%%%%%%%%%%%%%%%%%%%%%%%%%%%%%%%%%%%%%%%%%%%% FIG 1 %%%%%%%%%%%%%%%%%%%%%%%%%%%%%%%%%%%%%%%%%%%%%%%%%%%%%%%%%%%%%%%%%%%%%%
%%%%%%%%%%%%%%%%%%%%%%%%%%%%%%%%%%%%%%%%%%%%%%%%%%%%%%%%%%%%%%%%%%%%%%%%%%%%%%%%%%%%%%%%%%%%%%%%%%%%%%%%%%%%%%%%%%%%%%%%%%%%%%%%%%%%%%%%%%%%%%%%%%%
%%%%%%%%%%%%%%%%%%%%%%%%%%%%%%%%%%%%%%%%%%%%%%%%%%%%%%%%%%%%%%%%%%%%%%%%%%%%%%%%%%%%%%%%%%%%%%%%%%%%%%%%%%%%%%%%%%%%%%%%%%%%%%%%%%%%%%%%%%%%%%%%%%%
%%%%%%%%%%%%%%%%%%%%%%%%%%%%%%%%%%%%%%%%%%%%%%%%%%%%%%%%%%%%%%%%%%%% SECTION 2 %%%%%%%%%%%%%%%%%%%%%%%%%%%%%%%%%%%%%%%%%%%%%%%%%%%%%%%%%%%%%%%%%%%%
%%%%%%%%%%%%%%%%%%%%%%%%%%%%%%%%%%%%%%%%%%%%%%%%%%%%%%%%%%%%%%%%%%%%%%%%%%%%%%%%%%%%%%%%%%%%%%%%%%%%%%%%%%%%%%%%%%%%%%%%%%%%%%%%%%%%%%%%%%%%%%%%%%%
\section{Energy band diagram and Berry curvature}
\label{Energy diagram and Berry curvature}
% \begin{itemize}
%     \item The section title should be modified `later'.
%     \item In this section I will plot $E(k)$ diagram, Berry curvature of each bands, and It posses Chern number does not vary by magnetic field
% \end{itemize}
We study the magnon band structure of the negative vector chirality Kagome lattice under an external magnetic field. To perform the Holstein-Primakoff (HP) transformation, we introduce local coordinate systems whose $z$ axes are aligned with the classical spin directions at each atomic sublattice:
\begin{align}
    &\hat{S}^{x_0}_i=\tilde{S}^{x}_i\cos{\theta_i}-\tilde{S}^{y}_i\sin{\theta_i}\sin{\eta}-\tilde{S}^{z}_i\sin{\theta_i}\cos{\eta}, \\ \notag
    &\hat{S}^{y_0}_i=\tilde{S}^{x}_i\sin{\theta_i}+\tilde{S}^{y}_i\cos{\theta_i}\sin{\eta}+\tilde{S}^{z}_i\cos{\theta_i}\cos{\eta}, \\
    &\hat{S}^{z_0}_i=-\tilde{S}^{y}_i\cos{\eta}+\tilde{S}^{z}_i\sin{\eta}, \notag
\end{align}
where $\theta_i$ denotes sublattice-dependent angles with $\theta_\alpha = 4\pi/3, \theta_\beta = 2\pi/3, \theta_\gamma = 0$.
The non-interacting magnon Hamiltonian is obtained by applying the HP transformation in the linear spin wave approximation to the local spin operators, $\tilde{S}^{+}=\sqrt{2S}\hat{a},\, \tilde{S}^{-}=\sqrt{2S}\hat{a}^{\dagger},$ and $\tilde{S}^{z}= S-\hat{a}^{\dagger}\hat{a}$. We then perform a Fourier transformation from real space to momentum space using the convention, $\hat{a}_{i,l}=\frac{1}{\sqrt{N}}\sum_{\mathbf{k}}\hat{a}_{i,\mathbf{k}}e^{i\mathbf{k}\cdot(\mathbf{R}_l+\mathbf{r}_i)}$, where $\mathbf{R}_l$ is the Bravais lattice vector of the $l$th unit cell and $\mathbf{r}_i$ is the position of sublattice $i$ within the unit cell~\cite{holstein1940field}. The Hamiltonian [Eq.~(\ref{eq:Hamiltonian})] is then transformed into the following bosonic Bogoliubov-de-Gennes (BdG) form,
\begin{align}
    \hat{H}_2=\frac{1}{2}\sum_{\mathbf{k}}\hat{\psi}_{\mathbf{k}}^{\dagger}\hat{H}_{\mathbf{k}}\hat{\psi}_{\mathbf{k}},
    \label{eq:BdG Hamiltonian}
\end{align}
where $H_2$ denotes the quadratic Hamiltonian of the non-interacting magnon,
\begin{equation}
\hat{\psi}_\mathbf{k}=\left(\hat{a}_{1,\mathbf{k}},\cdots,\hat{a}_{\tau,\mathbf{k}},\hat{a}_{1,-\mathbf{k}}^{\dagger},\cdots,\hat{a}_{\tau,-\mathbf{k}}^{\dagger},\right)^{T}    
\end{equation}
is the Nambu spinor of the HP boson, and $\tau$ is sub-lattice index which is $\tau =3$ for our model. $\hat{H}_{\mathbf{k}}$ contains not only number-conserving terms in the diagonal $\tau\times\tau$ block, but also number-nonconserving terms in the off-diagonal $\tau\times\tau$ block. These include both intra- and inter-sublattice pairing terms, such as $\hat{a}_{1,\mathbf{k}}\hat{a}_{1,-\mathbf{k}}$ and $\hat{a}_{2,\mathbf{k}}\hat{a}_{3,-\mathbf{k}}$, which are characteristic of antiferromagnetic magnons~\cite{kamra2019antiferromagnetic,kamra2020magnon,lee2025diverging}. All such pairing terms contribute to shaping the quantum geometry of the magnon bands~\cite{lee2026berry}.

We obtain right-eigenvector $\left|u^n_\mathbf{k}\right>$ by diaginalizing the pseudo-Hermitian Hamiltonian, $\sigma_3 \hat{H}_{\mathbf{k}}$~\cite{colpa1978diagonalization,shindou2013topological,lee2018magnonic},
\begin{align}\label{eq:Hamiltonian-energy-solution}
    \sigma_3 \hat{H}_{\mathbf{k}}\left|u^n_\mathbf{k}\right>=\bar{\epsilon}_{n,\mathbf{k}}\left|u^n_\mathbf{k}\right>,
\end{align}
where $\bar{\epsilon}_{n,\mathbf{k}}$ is the pseudo-eigenenergy, which is $\bar{\epsilon}_{n,\mathbf{k}}=(\sigma_3)_{nn}\epsilon_{n,\mathbf{k}}$ with eigenenergy $\epsilon_{n,\mathbf{k}}$, $n$ is the band index, and $\sigma_3$ is the $2\tau\times 2\tau$ third Pauli matrix acting on the particle-hole space such that $\left< u^n_\mathbf{k}\right|\sigma_3 \left|u^m_\mathbf{k}\right>=(\sigma_3)_{nm}$. With this convention, the eigenvalue equation can be equivalently be written as $\bra{u_\mathbf{k}^n}\hat{H}_\mathbf{k} = \bar{\epsilon}_{n,\mathbf{k}}\bra{u_\mathbf{k}^n}\sigma_3$. In addition, the eigenvectors satisfy the completeness relation
\begin{align}\label{eq:Identity operator}
    I=\sum_{q}(\sigma_3)_{qq}\ket{u_\mathbf{k}^q}\bra{u_\mathbf{k}^q}\sigma_3.
\end{align}
The obtained energy bands are shown in Fig.~\ref{fig1}, where solid (dashed) lines represent bands without (with) the external magnetic field $B_{z_0}$. Note that $B_{z_0}$ makes the excitation energy of the lowest band (green) finite at the $\Gamma$ point and lift the degeneracy between the middle (orange) and the highest (purple) bands at the $\Gamma$ point. 

\begin{figure}
    \centering
    \includegraphics[width=\linewidth]{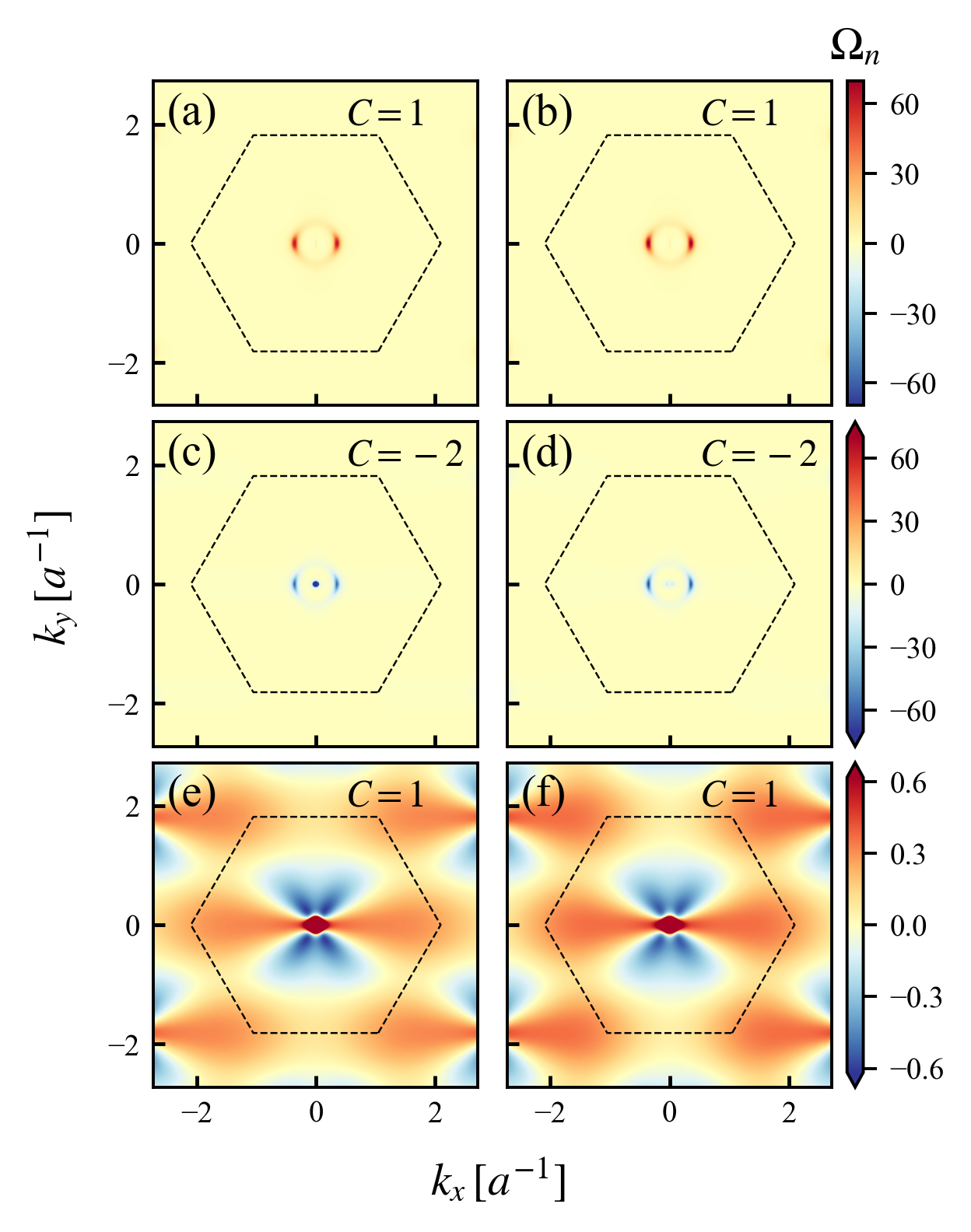}%임시로 png 사용
    \caption{Berry curvatures. Left (right) column: magnon Berry curvature texture in the negative vector chirality Kagome lattice at $0.05\,\text{T}$ ($1\,\text{T}$). Each row shows, from top to bottom, the Berry curvature of the [(a),(b)] lowest, [(c),(d)] middle band, and highest band [(e),(f)], respectively. Panels (c) and [(e),(f)] exhibit large negative (positive) Berry curvature values at the $\Gamma$ point. $\mathcal{C}$ denotes the Chern number for each bands.}
    % \textcolor{red}{1. n-th row means n-bands. 2. first (second) column means $0.05\,\mbox{T}$ $(1\,\mbox{T})$ picture. 3. BC dose not vary much by the external magnetic field. 4. Chern number are not vary by the magnetic field until $5\mbox{ T}$because it doesn't close the gap. 5. At $\Gamma$ point, the BC shows divergence picture thus I cut it off. 6. The positive (negative) divergence is occurs.}
    \label{fig2}
\end{figure}

We investigate the Berry curvature, which for the bosonic BdG Hamiltonian can be calculated from the following expression~\cite{xiao2010berry,go2024magnon}:
\begin{align}\label{eq:definition of BC}
     \Omega_{n,\mathbf{k}}=\frac{\partial A^{n}_{\mathbf{k},y}}{\partial k_x}-\frac{\partial A^{n}_{\mathbf{k},x}}{\partial k_y},\;\mathbf{A}^{n}_\mathbf{k}=\frac{\bra{u_\mathbf{k}^n}i\sigma_3\partial_{\mathbf{k}} \ket{u_\mathbf{k}^n}}{\bra{u_\mathbf{k}^n}\sigma_3 \ket{u_\mathbf{k}^n}}.
\end{align}
To obtain an interband representation of the Berry curvature, we choose the parallel-transport gauge, $\langle u^n_{\mathbf{k}}|\sigma_3|\partial_{k_i}u^n_{\mathbf{k}}\rangle=0$. By differentiating Eq.~(\ref{eq:Hamiltonian-energy-solution}) and using the completeness relation in Eq.~(\ref{eq:Identity operator}), we obtain
\begin{align}\label{eq:ket_derivative}
    \ket{\partial_{\mathbf{k}}u_{\mathbf{k}}^n}=\hbar\sum_{q\neq n}(\sigma_3)_{qq}\frac{\left<u_{\mathbf{k}}^q\middle|\hat{\mathbf{v}}\middle|u_{\mathbf{k}}^n\right>}{\bar{\epsilon}_{n,\mathbf{k}}-\bar{\epsilon}_{q,\mathbf{k}}}\ket{u_{\mathbf{k}}^q},
\end{align}
where $\hat{v}_i=(1/\hbar)\partial_{k_i}\hat{H}_{\mathbf{k}}$ is the velocity operator. Substituting this relation into Eq.~(\ref{eq:definition of BC}), the Berry curvature can be written in the Kubo-like interband form
\begin{align}
    \Omega_{n,\mathbf{k}} =& i\hbar^{2}\sum_{m\neq n}(\sigma_3)_{nn}(\sigma_3)_{mm}\frac{1}{(\bar{\epsilon}_{m,\mathbf{k}}-\bar{\epsilon}_{n,\mathbf{k}})^{2}} \notag \\
    &\quad\times\left( \left<u_{\mathbf{k}}^{n}\middle|\hat{v}_{x}\middle|u_{\mathbf{k}}^{m}\right>\left<u_{\mathbf{k}}^{m}\middle|\hat{v}_{y}\middle|u_{\mathbf{k}}^{m}\right>  - (k_x \leftrightarrow k_y)\right).
\end{align}
The Berry curvature texture and the Chern number, calculated under a finite external field, are shown in Fig~\ref{fig2}. The Chern numbers are $(1, -2, 1)$ from the lowest to the highest band and remain unchanged throughout the field range $0.05\,\text{T}$ to $5\,\text{T}$, as the band gaps do not close on this range (not shown). The energy bands and Berry curvatures are even in momentum $[\epsilon_{n,-\mathbf{k}} =\epsilon_{n,\mathbf{k}},\,\Omega_{n,-\mathbf{k}}= \Omega_{n,\mathbf{k}}]$, which allows for a non-zero thermal Hall effect~\cite{matsumoto2011theoretical, mook2019thermal}.
%In the field range from $0.05\,\text{T}$ to $5\,\text{T}$, the Chern numbers are $(1, -2, 1)$ from the lowest to the highest band and remain unchanged (not shown).%   %The energy bands and Berry curvatures are even in momentum $(\epsilon_\mathbf{-k} =\epsilon_\mathbf{k},\,\Omega_n(-\mathbf{k})= \Omega_n(\mathbf{k}))$, which results in a non-zero thermal Hall effect~\cite{mook2019thermal}.
%We diagonalize the BdG Hamiltonian by introducing the paraunitary matrix, $T_\mathbf{k}$, such that $T_\mathbf{k}^{\dagger}\sigma_3 T_\mathbf{k}=T_\mathbf{k}\sigma_3 T_\mathbf{k}^{\dagger}=\sigma_3$ where $\sigma_3$ is the $2N\times 2N$ third Pauli matrix for particle-hole space. The paraunitary matrix $T_\mathbf{k}$

%%%%%%%%%%%%%%%%%%%%%%%%%%%%%%%%%%%%%%%%%%%%%%%%%%%%%%%%%%%%%%%%%%%%%%%%%%%%%%%%%%%%%%%%%%%%%%%%%%%%%%%%%%%%%%%%%%%%%%%%%%%%%%%%%%%%%%%%%%%%%%%%%%%
%%%%%%%%%%%%%%%%%%%%%%%%%%%%%%%%%%%%%%%%%%%%%%%%%%%%%%%%%%%%%%%%%%%%%% FIG 2 %%%%%%%%%%%%%%%%%%%%%%%%%%%%%%%%%%%%%%%%%%%%%%%%%%%%%%%%%%%%%%%%%%%%%%
%%%%%%%%%%%%%%%%%%%%%%%%%%%%%%%%%%%%%%%%%%%%%%%%%%%%%%%%%%%%%%%%%%%%%%%%%%%%%%%%%%%%%%%%%%%%%%%%%%%%%%%%%%%%%%%%%%%%%%%%%%%%%%%%%%%%%%%%%%%%%%%%%%%
% \begin{itemize}
%     \item The section title should be modified `later'.
%     \item In this section, I will introduce a more detail definition and brief physical meaning of the total magnetic moment(TMM) and the orbital magnetic moment(OMM), especially the negative vector chirality Kagome lattice.
%     \item That is shown TMM and OMM are the same in zero magnetic field.
%     \item Does the multiple contour plot to be shown? or does one plot enough?
% \end{itemize}
%%%%%%%%%%%%%%%%%%%%%%%%%%%%%%%%%%%%%%%%%%%%%%%%%%%%%%%%%%%%%%%%%%%%%%%%%%%%%%%%%%%%%%%%%%%%%%%%%%%%%%%%%%%%%%%%%%%%%%%%%%%%%%%%%%%%%%%%%%%%%%%%%%%
%%%%%%%%%%%%%%%%%%%%%%%%%%%%%%%%%%%%%%%%%%%%%%%%%%%%%%%%%%%%%%%%%%%%%% FIG 3 %%%%%%%%%%%%%%%%%%%%%%%%%%%%%%%%%%%%%%%%%%%%%%%%%%%%%%%%%%%%%%%%%%%%%%
%%%%%%%%%%%%%%%%%%%%%%%%%%%%%%%%%%%%%%%%%%%%%%%%%%%%%%%%%%%%%%%%%%%%%%%%%%%%%%%%%%%%%%%%%%%%%%%%%%%%%%%%%%%%%%%%%%%%%%%%%%%%%%%%%%%%%%%%%%%%%%%%%%%
\begin{figure}
    \centering
    \includegraphics[width=\linewidth]{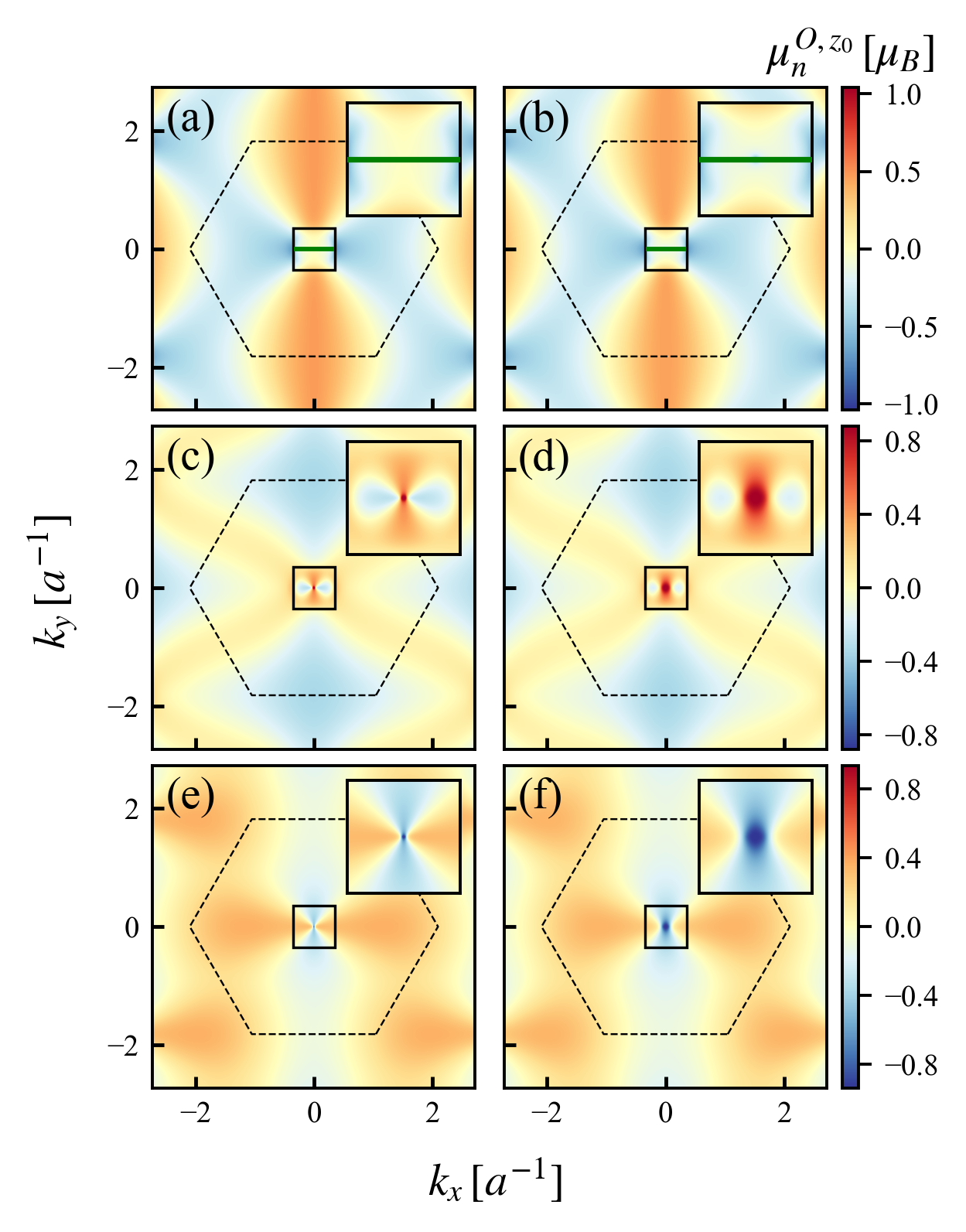}
    \includegraphics[width=\linewidth]{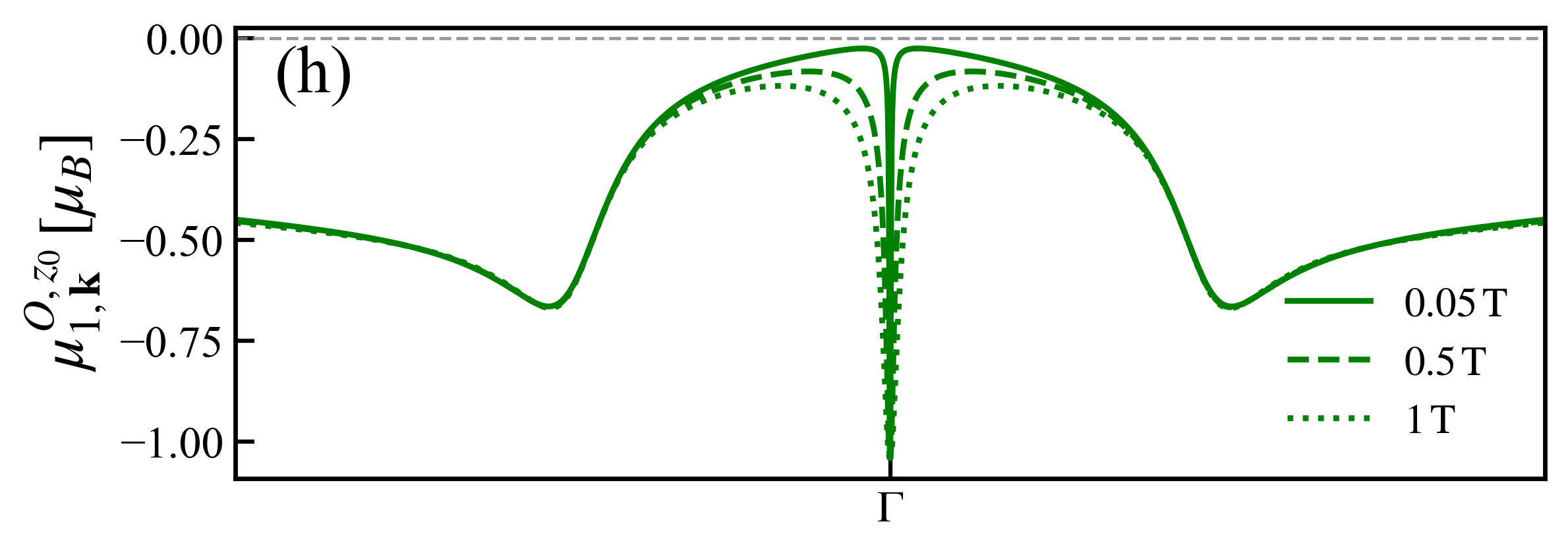}
    \caption{%Magnon OMM textures. Left (right) column: magnon OMM in the negative vector chirality Kagome lattice at $0.05\,\text{T}$ ($1\,\text{T}$). Each row shows, from top to bottom, the OMM of the (a),(b) lowest, (c),(d) middle band, and highest band (e),(f), respectively. The OMM values increase in magnitude near the $\Gamma$ point as the magnetic field increases.
    Magnon OMM textures. The left and right columns show the magnon OMM in the negative vector chirality Kagome lattice at $0.05\,\text{T}$ and $1\,\text{T}$, respectively. Each row corresponds, from top to bottom, to the OMM of the lowest band [(a), (b)], middle band [(c), (d)], and highest band [(e), (f)]. The inset in each panel shows a magnified view of the region around $\Gamma$ point enclosed by black square. Panel (h) show the lowest band OMM values plotted along the green lines indicated in panel (a) and (b). The solid, dashed, and dotted lines correspond to $B_{z_0}=0.05\,\mathrm{T},\,0.5\,\mathrm{T},$ and $\,1\,\mathrm{T},$ respectively.}
    % \textcolor{red}{1. n-th row means n-band. 2. first (second) column means $0.05\,\mbox{T}$ $(1\,\mbox{T})$ picture. 3. OMM change near the $\Gamma$ point much. 4. The definition of OMM must be written. 5. it's sum is shown in fig.5.}
    \label{fig3}
\end{figure}
%%%%%%%%%%%%%%%%%%%%%%%%%%%%%%%%%%%%%%%%%%%%%%%%%%%%%%%%%%%%%%%%%%%%%%%%%%%%%%%%%%%%%%%%%%%%%%%%%%%%%%%%%%%%%%%%%%%%%%%%%%%%%%%%%%%%%%%%%%%%%%%%%%%
%%%%%%%%%%%%%%%%%%%%%%%%%%%%%%%%%%%%%%%%%%%%%%%%%%%%%%%%%%%%%%%%%%%%%% FIG 4 %%%%%%%%%%%%%%%%%%%%%%%%%%%%%%%%%%%%%%%%%%%%%%%%%%%%%%%%%%%%%%%%%%%%%%
%%%%%%%%%%%%%%%%%%%%%%%%%%%%%%%%%%%%%%%%%%%%%%%%%%%%%%%%%%%%%%%%%%%%%%%%%%%%%%%%%%%%%%%%%%%%%%%%%%%%%%%%%%%%%%%%%%%%%%%%%%%%%%%%%%%%%%%%%%%%%%%%%%%
\begin{figure}
    \centering
    \includegraphics[width=\linewidth]{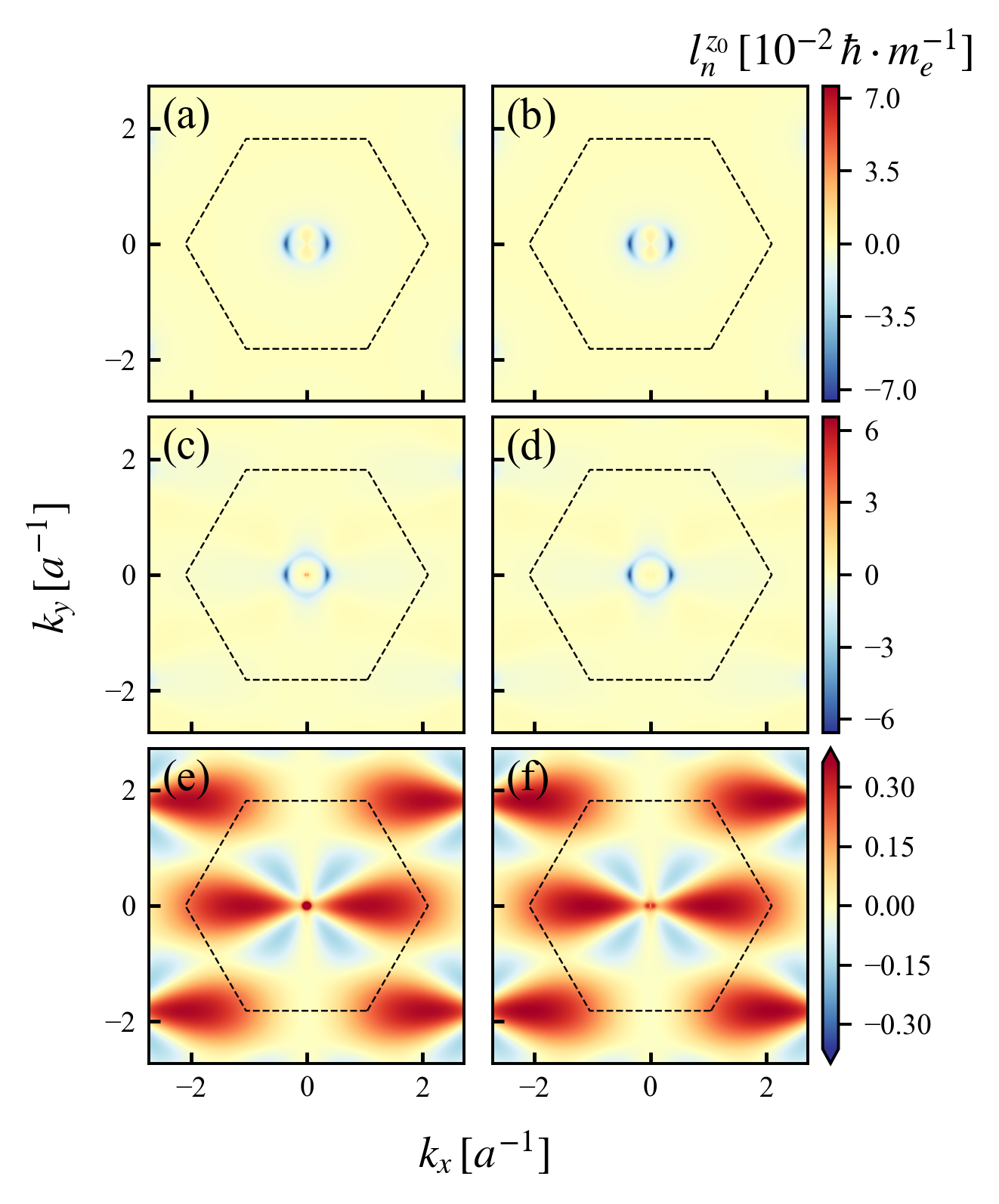}%임시로 png 사용
    \caption{%Magnon OAM textures. Left (right) column: magnon OAM in the negative vector chirality Kagome lattice at $0.05\,\text{T}$ ($1\,\text{T}$). Each row shows, from top to bottom, the OAM of the (a),(b) lowest, (c),(d) middle band, and highest band (e),(f), respectively. 
    Magnon OAM textures for each band in the negative vector chirality Kagome lattice. The left and right panels show the magnon OAM distributions at $0.05\,\text{T}$ and $1\,\text{T}$, respectively. Panels (a) and (b) correspond to the lowest band, (c) and (d) to the middle band, and (e) and (f) to the highest band.}
    %\textcolor{red}{1. n-th row means n-band. 2. first (second) column means $0.05\,\mbox{T}$ $(1\,\mbox{T})$ picture. 3. OAM dose not vary much by the external magnetic field like BC. 4. it's sum also does not vary much than OMM in comparison that shown in fig .5. 5. Well-matched to Neuman's one. 6. Even in $\mathbf{k}$ but why is it so?}}
    \label{fig4}
\end{figure}
\section{Magnon orbital magnetic moment}
\label{Magnon orbital magnetic moment}
The magnetic moment, $\mu_{n,\mathbf{k}}^{z_0}$, defined as the derivative of energy with respect to the magnetic field, consists not only of the conventional spin magnetic moment, $\mu_{n,\mathbf{k}}^{S,z_0}$, but also an additional OMM, $\mu_{n,\mathbf{k}}^{O,z_0}$~\cite{neumann2020orbital},
\begin{align}\label{eq:intraband OMM}
    \mu^{z_0}_{n,\mathbf{k}} &=-\frac{\partial\epsilon_{n,\mathbf{k}}}{\partial B_{z_0}} \\
    &= -\left< u_{\mathbf{k}}^{n}\middle| \frac{\partial \hat{H}_\mathbf{k}}{\partial B_{z_0}} \middle| u_{\mathbf{k}}^{n}\right> \notag \\
    & = -\left< u_{\mathbf{k}}^{n}\middle| \frac{\partial \hat{H}_\mathbf{k}}{\partial B_{z_0}} - g\mu_B \hat{S}^{z_0}_\mathbf{k} \middle| u_{\mathbf{k}}^{n}\right>-\left< u_{\mathbf{k}}^{n}\middle| g\mu_B \hat{S}^{z_0}_\mathbf{k} \middle| u_{\mathbf{k}}^{n}\right> \notag \\ 
    &= \mu^{O,z_0}_{n,\mathbf{k}} + \mu^{S,z_0}_{n,\mathbf{k}}. \notag
\end{align}
In this paper, we only deal with the $z_0$-component of the OMM of magnons, i.e., $\mu_{n,\mathbf{k}}^{O,z_0}=-\partial\epsilon_{n,\mathbf{k}}/\partial B_{z_0}+g\mu_B S^{z_{0}}_{n,\mathbf{k}},$ where $S^{z_0}_{n,\mathbf{k}}$ is the $z_0$-component of the magnon spin expectation value for $n$-th band. This definition allows us to isolate the orbital contribution to the magnetic moment and analyze its behavior independently of the spin sector. The magnon OMM textures calculated for each band at a fixed magnetic field ($ 0.05\mbox{ T }$ and $1\mbox{ T}$) are shown in Fig.~\ref{fig3} and are found to be in good agreement with previous results~\cite{neumann2020orbital}, validating our implementation.

The total orbital magnetization, $\left<\mu_{\mathrm{tot}}^{O,z_0}\right>$, obtained by incorporating the Bose-Einstein distribution, is given by
\begin{align}
    \left<\mu_{\mathrm{tot}}^{O,z_0}\right>(B_{z_0}, T)=\frac{1}{V}\sum_{n,\mathbf{k}}\mu_{n,\mathbf{k}}^{O,z_0}(B_{z_0})\,\rho(\epsilon_{n,\mathbf{k}}(B_{z_0}),T),
\end{align}
where $\rho(\epsilon)=(\exp{(\epsilon/k_{\mathrm{B}}T)-1})^{-1}$ with Boltzmann constant, $k_{\mathrm{B}}$, is the Bose-Einstein distribution. The resulting total orbital magnetization is presented in Fig.~\ref{fig5}(a) as red curves. The OMM textures are even in momentum ($\mu_{n,-\mathbf{k}}^{O,z_0}=\mu_{n,\mathbf{k}}^{O,z_0}$) regardless of the magnetic field strength. As the field increases, the OMM becomes more pronounced at the $\Gamma$ point (Fig.~\ref{fig3}), where most of the change originates; this trend is also reflected in the orbital magnetization, which shifts from negative to positive values with increasing magnetic field strength [Fig.~\ref{fig5}(a)].
\section{Magnon orbital angular momentum}
\label{Magnon orbital angular momentum}
Like electrons, magnons can also possess an OAM arising from the itinerant motion of their wave packets. In electronic systems, this OAM contributes to the orbital magnetization and plays a crucial role in transport phenomena~\cite{go2018intrinsic, choi2023observation}. Recent theoretical studies have extended this concept to magnons, demonstrating that magnon OAM and its transport properties~\cite{fishman2022orbital, fishman2022exact, fishman2023gauge, go2024magnon} are likewise significant. To capture the OAM in our system, we adopt the definition formulated in Refs.~\cite {xiao2005berry, thonhauser2005orbital, shi2007quantum}. We start with symmetrized OAM operator, which is the following~\cite{bhowal2021orbital,pezo2022orbital}:
\begin{align}\label{OAM operator}
    \hat{\mathbf{l}}=\frac{1}{4}(\hat{\mathbf{r}}\times\hat{\mathbf{v}}-\hat{\mathbf{v}}\times\hat{\mathbf{r}}).
\end{align}
Then, the matrix element of the $a$-component of the OAM is definded as
\begin{align}\label{eq:interband OAM_1}
    l^{(a)}_{nm,\mathbf{k}}&=\frac{\varepsilon_{abc}}{4}\bra{u_\mathbf{k}^n}(\hat{r}_b \hat{v}_c-\hat{v}_b \hat{r}_c)\ket{u_\mathbf{k}^{m}} \\
    &=\frac{\varepsilon_{abc}}{4i\hbar}\bra{u_\mathbf{k}^n}(\hat{r}_b[\hat{r}_c,\hat{H}_\mathbf{k}]-[\hat{r}_b,\hat{H}_\mathbf{k}]\hat{r}_c)\ket{u_\mathbf{k}^m}.
\end{align}
Here we use the relation $\hat v_i=(1/i\hbar)[\hat r_i,\hat{H}_{\mathbf{k}}]$. In momentum space, the position operator is represented by $\hat r_i=i\partial_{k_i}$ when acting on the $\mathbf{k}$-dependent eigenstates. Using this relation, the above expression can be rearranged as

\begin{align}\label{eq:interband OAM_2}
    l^{(a)}_{nm,\mathbf{k}}=&-\frac{\varepsilon_{abc}}{2i\hbar}\bra{\partial_{k_b}u_\mathbf{k}^{n}}\hat{H}_\mathbf{k}\ket{\partial_{k_c}u_\mathbf{k}^{m}} \\ \notag
    &+\frac{\varepsilon_{abc}}{4i\hbar}\bra{u_\mathbf{k}^n}\hat{r}_{b}\hat{r}_{c}\sigma_3\sigma_3\hat{H}_\mathbf{k}+\hat{H}_\mathbf{k}\hat{r}_{b}\hat{r}_{c}\ket{u_\mathbf{k}^{m}}.
\end{align}
The second term can be simplified by inserting $\sigma_3\sigma_3=I$ and using the bosonic BdG eigenvalue eqation,
\begin{align}\label{eq:interband OAM_3}
    l^{(a)}_{nm,\mathbf{k}}=&-\frac{\varepsilon_{abc}}{2 i\hbar}\bra{\partial_{k_b}u_\mathbf{k}^n}\hat{H}_\mathbf{k}\ket{\partial_{k_c}u_\mathbf{k}^m}\\ \notag
    &+\frac{\varepsilon_{abc}}{2 i\hbar}\frac{\bar{\epsilon}_{n,\mathbf{k}}+\bar{\epsilon}_{m,\mathbf{k}}}{2}\bra{\partial_{k_b}u_\mathbf{k}^n}\sigma_{3}\ket{\partial_{k_c}u_\mathbf{k}^m}.
\end{align}
For the diagonal element, $m=n$, this expression reduces to the band-resolved OAM,
\begin{align}\label{eq:intraband OAM_1}
    l^{(a)}_{n,\mathbf{k}}&=-\frac{\varepsilon_{abc}}{2 i\hbar}\bra{\partial_{k_b}u_\mathbf{k}^n}(\hat{H}_\mathbf{k}-\sigma_3 \bar{\epsilon}_{n,\mathbf{k}})\ket{\partial_{k_c}u_\mathbf{k}^n},
\end{align}
For the transport calculation, we also need the off-diagonal OAM matrix elements. Using Eq.~(\ref{eq:ket_derivative}), the matrix element can be written as
\begin{align}\label{eq:interband OAM numerator form}
    l^{(a)}_{nm,\mathbf{k}}=&i\frac{\hbar}{4}\sum_{q\neq n,m}(\sigma_3)_{qq}\left(\frac{1}{\bar{\epsilon}_{q,\mathbf{k}}-\bar{\epsilon}_{n,\mathbf{k}}}+\frac{1}{\bar{\epsilon}_{q,\mathbf{k}}-\bar{\epsilon}_{m,\mathbf{k}}}\right)\notag \\ 
    &\quad\times\varepsilon_{abc}\left<u_{\mathbf{k}}^{n}\middle|\hat{v}_{b}\middle|u_{\mathbf{k}}^{q}\right>\left<u_{\mathbf{k}}^{q}\middle|\hat{v}_{c}\middle|u_{\mathbf{k}}^{m}\right>.
\end{align}
This final expression makes it explicit that the OAM originates from interband virtual processes mediated by intermediate magnon bands $q$.
Here $\varepsilon_{abc}$ is the Levi-Civita symbol, and $a,b,c$ run over $x_0,y_0,z_0$. The calculated $z_0$-component of OAM textures is shown in Fig.~\ref{fig4}. Similar to the Berry curvature, the OAM exhibits little variation under the applied magnetic field and is even in momentum ($l_{n,-\mathbf{k}}^{z_0}=l_{n,\mathbf{k}}^{z_0}$). As shown in Fig.~\ref{fig4}, the OAM texture closely resembles the magnon Berry curvature, indicating that the OAM, like the Berry curvature, originates from interband processes~\cite{lee2025universal}. In contrast, the magnon OMM texture (Fig.~\ref{fig3}) exhibits a qualitatively different structure from both the Berry curvature (Fig.~\ref{fig2}) and the magnon OAM textures (Fig.~\ref{fig4}). In particular, its response to an external magnetic field is highly sensitive near the $\Gamma$ point. This behavior reflects the distinct interband nature of the magnon OMM.

This behavior is also reflected in the total average OAM $\left<l_{\mathrm{tot}}^{z_0}\right>$, shown as blue curves in Fig.~\ref{fig5}(a), which is obtained by weighting the OAM with the Bose–Einstein distribution,
\begin{align}
    \left<l_\mathrm{tot}^{z_0}\right>(B_{z_0},T) = \frac{1}{V}\sum_{n,\mathbf{k}}l_{n,\mathbf{k}}^{z_0}(B_{z_0})\,\rho(\epsilon_{n,\mathbf{k}}(B_{z_0}),T).
\end{align}
As shown in Fig.~\ref{fig5}(a), while the OMM exhibits a relatively strong response to the magnetic field the average value of the OAM is only weakly affected, highlighting the contrasting behaviors of the two quantities.
% The magnon orbital angular moment (OAM) is the self-rotation of the magnon wave-packet~\cite{xiao2010berry,fishman2022orbital,go2024magnon}. In this work, we follow the definition of the OAM as formulated in Ref.~\cite{xiao2005berry, thonhauser2005orbital, shi2007quantum}. The OAM 
% \begin{itemize}
%     \item{Magnon spin texture}\\
%     Magnon spin texture with or without $D_p$ and its proportion at a high symmetry point or high-value point 
%     \item{Magnon orbital texture}\\
%     Magnon orbital texture with or without $D_p$ and its proportion at a high symmetry point or high-value point
% \end{itemize}

%%%%%%%%%%%%%%%%%%%%%%%%%%%%%%%%%%%%%%%%%%%%%%%%%%%%%%%%%%%%%%%%%%%%%%%%%%%%%%%%%%%%%%%%%%%%%%%%%%%%%%%%%%%%%%%%%%%%%%%%%%%%%%%%%%%%%%%%%%%%%%%%%%%
%%%%%%%%%%%%%%%%%%%%%%%%%%%%%%%%%%%%%%%%%%%%%%%%%%%%%%%%%%%%%%%%%%%%%%%%%%%%%%%%%%%%%%%%%%%%%%%%%%%%%%%%%%%%%%%%%%%%%%%%%%%%%%%%%%%%%%%%%%%%%%%%%%%
%%%%%%%%%%%%%%%%%%%%%%%%%%%%%%%%%%%%%%%%%%%%%%%%%%%%%%%%%%%%%%%%%%%%%%%%%%%%%%%%%%%%%%%%%%%%%%%%%%%%%%%%%%%%%%%%%%%%%%%%%%%%%%%%%%%%%%%%%%%%%%%%%%%
\section{Transport of the magnons}
\label{Transport of the magnon}
%%%%%%%%%%%%%%%%%%%%%%%%%%%%%%%%%%%%%%%%%%%%%%%%%%%%%%%%%%%%%%%%%%%%%%%%%%%%%%%%%%%%%%%%%%%%%%%%%%%%%%%%%%%%%%%%%%%%%%%%%%%%%%%%%%%%%%%%%%%%%%%%%%%
%%%%%%%%%%%%%%%%%%%%%%%%%%%%%%%%%%%%%%%%%%%%%%%%%%%%%%%%%%%%%%%%%%%%%% FIG 6 %%%%%%%%%%%%%%%%%%%%%%%%%%%%%%%%%%%%%%%%%%%%%%%%%%%%%%%%%%%%%%%%%%%%%%
%%%%%%%%%%%%%%%%%%%%%%%%%%%%%%%%%%%%%%%%%%%%%%%%%%%%%%%%%%%%%%%%%%%%%%%%%%%%%%%%%%%%%%%%%%%%%%%%%%%%%%%%%%%%%%%%%%%%%%%%%%%%%%%%%%%%%%%%%%%%%%%%%%%
\begin{figure}
    \centering
    \includegraphics[width=1\linewidth]{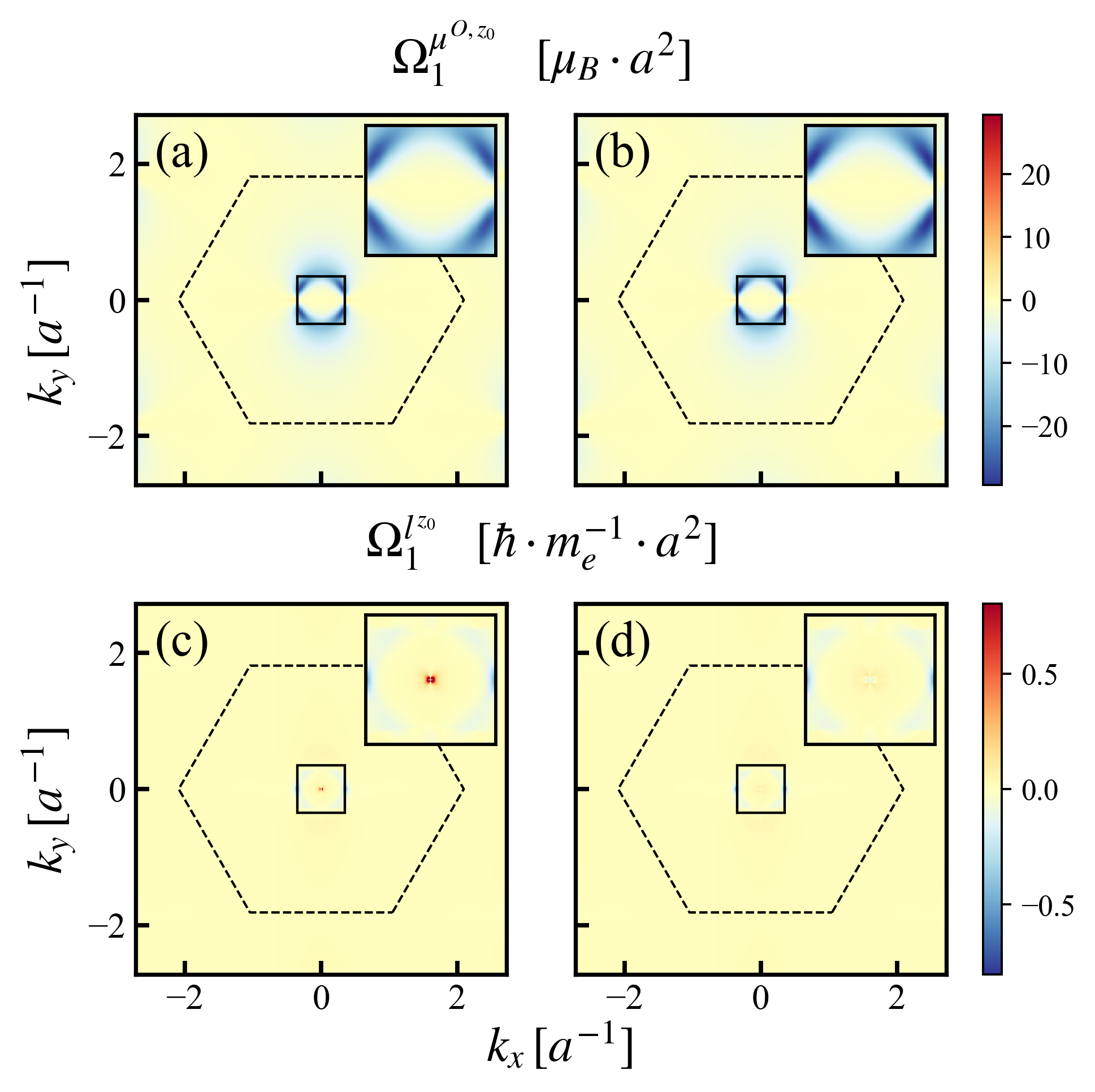}
    \caption{OMM and OAM Berry curvatures texture of the lowest band in momentum space. (a),(b) OMM Berry curvature $\Omega^{\mu^{O,z_0}}_{1}$ at $B_{z_0}=0.05~\mathrm{T}$ and $1~\mathrm{T}$, respectively. (c),(d) OAM Berry curvature $\Omega^{l^{z_0}}_{1}$ at $B_{z_0}=0.05~\mathrm{T}$ and $1~\mathrm{T}$, respectively. The inset in each panel shows a magnified view of the region around $\Gamma$ point enclosed by black square.}
    \label{fig:fig6}
\end{figure}
%\begin{itemize}
%    \item{Spin Nernst effect}\\
%    Magnon spin Nernst effect with or without $D_p$ and its proportion with $D_p$ and temperature, $T$.\\
%    \jl{[JL: We may emphasize it is the Nernst effect of the ``squeezed magnon" with non-integer spins.]}
%    \item{Orbital Nernst effect}\\
%    Magnon orbital Nernst effect with or without $D_p$ and its proportion with $D_p$ and temperature, $T$.
%\end{itemize}
Just as a temperature gradient gives rise to the magnon spin Nernst effect~\cite{zyuzin2016magnon, cheng2016spin, li2020intrinsic}, where magnon spin flows transversely to the gradient, OMM and OAM can also exhibit analogous transverse responses.  
This gives rise to the magnon OMM and OAM Nernst effects, in which the orbital degrees of freedom flow perpendicular to the thermal gradient. We calculate the Nernst coefficient for each observable $\mathcal{O}$, defined as the transverse orbital current of $z$-component $J^{\mathcal{O}}_{y}$ that flows in the $y$-direction under a temperature gradient applied along the $x$-direction i.e $J^{\mathcal{O}}_y=-\alpha^{\mathcal{O}}_z\partial_x T$. The $\mathcal{O}$-Nernst coefficient, $\alpha^{\mathcal{O}}_z$, is given by
\begin{align}
    \alpha_z^\mathcal{O}=\frac{2k_B}{\hbar V_{\mathrm{uc}}}\sum_{n,\mathbf{k}}c_1\left(\rho(\epsilon_{n,\mathbf{k}})\right)\Omega_{n,\mathbf{k}}^{\mathcal{O}},
\end{align}
where the weighting function, $c_1$ is defined as $c_1(\rho) = (1+\rho)\ln{(1+\rho)} - \rho\ln{(\rho)}$~\cite{matsumoto2011berry}, and the $\mathcal{O}$-Berry curvature is given by
\begin{align}\label{eq:O-BC}
    \Omega_{n,\mathbf{k}}^{\mathcal{O}}=2\hbar^2\sum_{m\neq n}G_{nm}
    \frac{\mathrm{Im}\left[\langle u_\mathbf{k}^n | \hat{j}^{\mathcal{O}}_{y} | u_\mathbf{k}^m \rangle
    \langle u_\mathbf{k}^m | \hat{v}_x | u_\mathbf{k}^n \rangle\right]}
    {(\bar{\epsilon}_{n,\mathbf{k}} - \bar{\epsilon}_{m,\mathbf{k}})^2},
\end{align}
where $G_{nm}= (\sigma_3)_{nn}(\sigma_3)_{mm}$. For bosonic BdG systems, the current operator is defined as $\hat{j}^{\mathcal{O}}_{y} = (\hat{v}_y \sigma_3 \mathcal{O} + \mathcal{O} \sigma_3 \hat{v}_y)/2 $~\cite{matsumoto2014thermal,zyuzin2016magnon,go2024magnon} where $\mathcal{O}$ denotes either the OMM operator,
\begin{align}\label{eq. 15}
    \mathcal{O}=-\frac{\partial \hat{H}_{\mathbf{k}}}{\partial B_{z_0}}+g\mu_B \hat{S}^{z_0}_{\mathbf{k}},
\end{align}
Eq.~(\ref{eq. 15}) rewrites Eq.~(\ref{eq:intraband OMM}), where only the $z_0$-component of the magnon OMM is considered, into an operator form that explicitly incorporates interband effects, or the OAM operator,
\begin{align}\label{eq. 16}
    \mathcal{O}=\left.\frac{1}{4}(\hat{\mathbf{r}}\times\hat{\mathbf{v}}-\hat{\mathbf{v}}\times\hat{\mathbf{r}})\right|_{z_0},
\end{align}
and Eq.~(\ref{eq. 16}) is used in Eq.~(\ref{eq:interband OAM_1}).

 The $\mathcal{O}$-Berry curvatures calculated at $B_{z_0}=0.05\,\mathrm{T}$ and $1\,\mathrm{T}$ are shown in Fig.~\ref{fig:fig6}. For the OMM Berry curvature, increasing the magnetic field changes mainly the overall magnitude of the texture, while leaving its momentum-space structure nearly unchanged. In contrast, the OAM Berry curvature develops a peak around the $\Gamma$ point at $B_{z_0}=0.05\,\mathrm{T}$, and the magnitude of this peak decreases as the magnetic field increases.

\section{Results and Discussion}
\label{Results and Discussion}
%%%%%%%%%%%%%%%%%%%%%%%%%%%%%%%%%%%%%%%%%%%%%%%%%%%%%%%%%%%%%%%%%%%%%%%%%%%%%%%%%%%%%%%%%%%%%%%%%%%%%%%%%%%%%%%%%%%%%%%%%%%%%%%%%%%%%%%%%%%%%%%%%%%
%%%%%%%%%%%%%%%%%%%%%%%%%%%%%%%%%%%%%%%%%%%%%%%%%%%%%%%%%%%%%%%%%%%%%% FIG 5 %%%%%%%%%%%%%%%%%%%%%%%%%%%%%%%%%%%%%%%%%%%%%%%%%%%%%%%%%%%%%%%%%%%%%%
%%%%%%%%%%%%%%%%%%%%%%%%%%%%%%%%%%%%%%%%%%%%%%%%%%%%%%%%%%%%%%%%%%%%%%%%%%%%%%%%%%%%%%%%%%%%%%%%%%%%%%%%%%%%%%%%%%%%%%%%%%%%%%%%%%%%%%%%%%%%%%%%%%%
\begin{figure}[t]
    \centering
    \includegraphics[width=\linewidth]{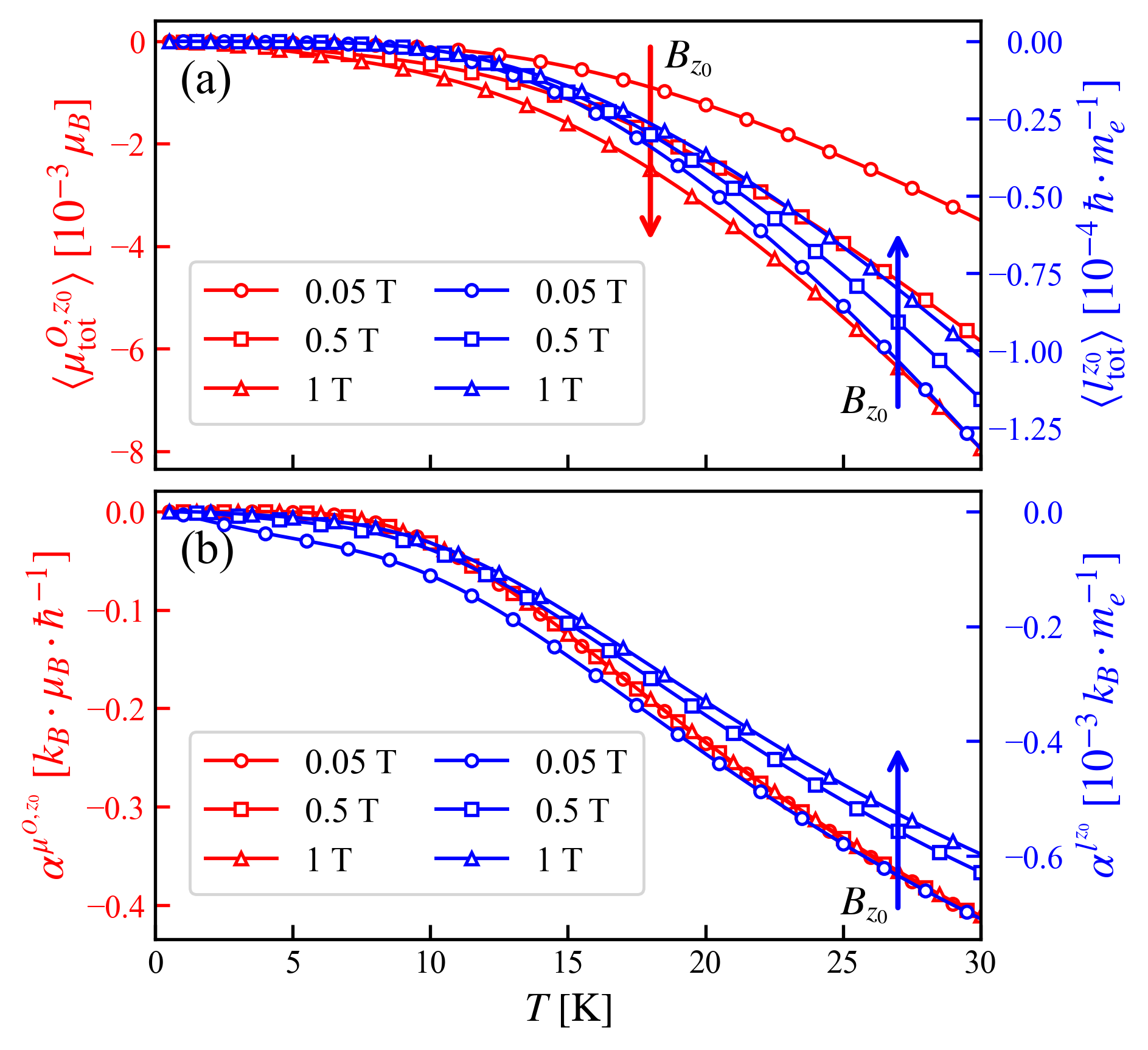}%임시로 png 사용
    \caption{(a) Temperature and magnetic field dependences of the total orbital magnetic moment $\left<\mu_{\mathrm{tot}}^{O,z_0}\right>$ (red) and the total magnon orbital angular momentum $\left< l_{\mathrm{tot}}^{z_0}\right>$ (blue) (b) Nernst coefficient $\alpha^{\mu_{\mathrm{tot}}^{O,z_0}}$ (red) of the total orbital magnetic moment and the Nernst coefficient $\alpha^{l_{\mathrm{tot}}^{z_0}}$ (blue) of the total angular momentum. The circle, square, and triangle markers correspond to $B_{z_0}=0.05$, $0.5$, and $1.0~{\rm T}$, respectively. In panel (b), the red curves for different magnetic fields almost completely overlap. The arrows indicate the direction in which the corresponding quantities change as $B_{z_0}$ increases, with the arrow colors matching the corresponding axes.}
    % (a) Temperature and magnetic field dependences of the total orbital magnetic moment $<\mu_{\mathrm{tot}}^{O,z}>$ (red) and the total magnon orbital angular momentum $l_{\mathrm{tot}}$ (blue) (b) Nernst coefficient <\alpha^{\mu_{\mathrm{tot}}^{O,z}}> (red) of the total orbital magnetic moment and the Nernst coefficient $\alpha^{l^{\mathrm{tot}}}$ (blue) of the total angular momentum
    %\textcolor{red}{1. (a),(c): it's average values are calculated with the Bose-Einstein distribution. 2. (a),(c): reds are change its sign by the external magnetic field while blues are not. it can be predicted in fig.4 and 5. 3. (b): reds's and blues's tendency looks very similar, but the sign while average values are different. 4. (d): reds are nearly constant while average are not. 5. (d): At low field, it's tendency looks different, but at high field regime it's tendency very similar.}}
    \label{fig5}
\end{figure}
Since the present Hamiltonian is intended as a minimal model rather than a material-specific description, we do not assign a specific Néel temperature to the system. The finite-temperature results should therefore be understood as lower-temperature trends within the linear spin-wave approximation, rather than as quantitative predictions for a particular material. In Fig.~\ref{fig5}, we restrict the plotted temperature range to $T\le 30\,\mathrm{K}$ in order to avoid emphasizing the high-temperature regime, while retaining sufficient temperature dependence to compare the equilibrium and transport responses of the OMM and OAM.

The thermodynamic averages of the OMM and OAM are shown in Fig.~\ref{fig5}(a). As the magnetic field becomes stronger, the magnitude of the OMM grows, whereas that of the OAM decreases only weakly. This strong field dependence of the OMM originates mainly from the pronounced peak that develops near the $\Gamma$ point, as shown in Fig.~\ref{fig3}.

Figure~\ref{fig5}(b) shows the Nernst coefficients associated with the OMM and OAM. In contrast to the thermodynamic averages in Fig.~\ref{fig5}(a), the two Nernst coefficients exhibit broadly similar temperature dependences. The OMM Nernst coefficient is nearly insensitive to the magnetic field, while the OAM Nernst coefficient approaches a field-independent behavior as the field strength increases.

The spin Nernst effect of magnons has been theoretically studied for $\mathrm{MnPS_3}$ in Refs.~\cite{zyuzin2016magnon,cheng2016spin}. In addition, an experiment on $\mathrm{Pt/MnPS_3}$ hybrid structures reported low-temperature thermoelectric voltage signals consistent with the magnon Nernst effect, with the induced inverse spin Hall voltage estimated to be at most several tens of nanovolts~\cite{shiomi2017experimental}. This experimentally measured voltage is not a direct measure of the spin Nernst coefficient itself, since it also depends on interfacial spin transmission and spin-to-charge conversion in $\mathrm{Pt}$. Nevertheless, it demonstrates that magnon Nernst-induced transverse responses in $\mathrm{MnPS_3}$-based systems can be experimentally detectable. In this context, the OMM Nernst coefficient obtained in our model at $T=30\,\mathrm{K}$ and $B_{z_0}=0.05\,\mathrm{T}$ is comparable in magnitude to the theoretically estimated spin Nernst coefficient of $\mathrm{MnPS_3}$, while our OAM Nernst coefficient is about one order of magnitude larger than the value reported in Ref.~\cite{go2024magnon}.

To further clarify the origin of the field dependence of the Nernst coefficients, we show in Fig.~\ref{fig:fig6} the OMM and OAM Berry curvatures of the lowest magnon band at $B_{z_0}=0.05$ and $1~{\rm T}$. The OMM Berry curvature shows only a weak change in its overall magnitude, with little modification of its momentum-space texture. The OAM Berry curvature also retains its overall texture, although it exhibits a peak near the $\Gamma$ point at weak magnetic field whose magnitude is reduced as the field increases. This behavior is consistent with the ordinary Berry curvature in Fig.~\ref{fig2} and the OAM texture in Fig.~\ref{fig4}, which are also relatively insensitive to the external field compared with the OMM texture.
Thus, at least for the present system, band geometry related quantities such as the Berry curvature [Eq.~(\ref{eq:definition of BC})], OAM [Eq.~(\ref{eq:intraband OAM_1})], and $\mathcal{O}$-Berry curvature [Eq.~(\ref{eq:O-BC})] are much less field-sensitive than the OMM itself. More specifically, as shown in Fig.\ref{fig:fig6}, the field-induced gap opening at the $\Gamma$ point is not appreciably reflected in the OMM Berry curvarture, whereas it is directly manifested in the OAM Berry curvature through the formation of a peak near the $\Gamma$ point.This explains why the OMM and OAM Nernst coefficients in Fig.~\ref{fig5}(b) show similar field dependences despite the markedly different equilibrium behaviors shown in Fig.~\ref{fig5}(a). The remaining field dependence of the OAM Nernst coefficient can be attributed to the Bose statistical weight, which emphasizes low-energy states near the $\Gamma$ point. As a result, the field-induced peak of the OAM Berry curvature near the $\Gamma$ point, and its reduction at higher fields, are reflected in the OAM Nernst response.

A similar contrast between the equilibrium and transport responses of the OMM and OAM may also arise in other magnonic systems that host nontrivial OMM and OAM textures. Such behavior may be expected when the external magnetic field modifies the OMM more strongly than the underlying band geometry, leaving quantities such as the Berry curvature, the OAM, and the $\mathcal{O}$-Berry curvature relatively insensitive to the field. However, the extent to which this behavior occurs may depend on the magnetic order, symmetry, and magnon band structure of the system. Further studies of different magnonic systems are therefore required to determine how broadly the behavior found here applies.

\section{Conclusion}
\label{Conclusion and discussion}
In this work, we conducted a detailed comparison between the OMM and OAM of magnons in a negative vector chirality Kagome antiferromagnet under an external magnetic field. 
By contrasting the thermodynamic definition of the OMM with the wave packet based definition of the OAM, we showed that these two quantities exhibit markedly different equilibrium behaviors.

 Despite the clear difference in their equilibrium responses, the OMM and OAM Nernst coefficients show similar temperature and magnetic-field dependences. This similarity arises because the transport response is governed mainly by the $\mathcal{O}$-Berry curvature, a band geometry related quantity that is relatively insensitive to the magnetic field in the present system. Taken together, these results show that equilibrium moments and transverse thermal transport reflect different aspects of magnon orbital dynamics. While the equilibrium OMM is sensitive to the field-induced modification near the $\Gamma$ point, the transport response is controlled more robustly by the geometric structure of the magnon bands.

Our results clarify the distinctions and similarity between the OMM and OAM of magnons as chargeless bosonic quasiparticles, and provide a understanding of orbital responses of magnons in Kagome lattices and other magnonic systems with nontrivial band geometry.

\acknowledgements
We acknowledge support from the National Research Foundation of Korea (NRF), funded by the Korean government (MSIT) (Grants No. RS-2024-00356270 and No. RS-2024-00410027). J.M.L. acknowledges support from Quantum Horizons Alberta. We thank Seungyun Han, Jeonghun Sohn, and Hojun Lee for valuable discussions.

\bibliography{BibRef}

\end{document}